\documentclass[a4paper,fleqn,usenatbib]{mnras}

\usepackage{newtxtext,newtxmath}
\usepackage[T1]{fontenc}
\usepackage{ae,aecompl}

\usepackage{graphicx}	% Including figure files
\usepackage{amsmath}	% Advanced maths commands
\usepackage{amssymb}	% Extra maths symbols
\usepackage{xspace}
\usepackage{url}
\usepackage{breakurl}

\newcommand{\logg}{\mbox{$\log g$}\xspace}
\newcommand{\loggw}[1]{\mbox{$\log g\hspace{-0.5mm} =\hspace{-0.5mm}  #1$}}
\newcommand{\Teff}{\mbox{$T_\mathrm{eff}$}\xspace}

\newcommand{\Teffw}[1]{\mbox{$\Teff\hspace{-0.5mm} =\hspace{-0.5mm} #1 \,\mathrm{K}$}}
\newcommand{\ebv}{\mbox{$E_\mathrm{B-V}$}}
\newcommand{\ebvw}[1]{\mbox{$\ebv\hspace{-0.5mm} =\hspace{-0.5mm} #1$}}

\newcounter{Rco}
\newcommand{\Ionst}[1]{\setcounter{Rco}{#1}\Roman{Rco}}
\newcommand{\Ion}[2]{\mbox{#1\,{\scriptsize\Ionst{#2}}}}
\newcommand{\Ionw}[3]{\mbox{#1\,{\scriptsize\Ionst{#2}}~$\lambda\,#3$\,\AA}\xspace}

\newcommand{\Ionww}[3]{\mbox{#1\,{\scriptsize\Ionst{#2}}~$\lambda\lambda\,#3$\,\AA}\xspace}
\newcommand{\ta}[1]{\mbox{Table~\ref{#1}}}
\newcommand{\sT}[1]{\mbox{(Table~\ref{#1})}}
\newcommand{\ab}[1]{\mbox{Fig.~\ref{#1}}}
\newcommand{\sA}[1]{\mbox{(Fig.~\ref{#1})}}
\newcommand{\se}[1]{\mbox{Sect.~\ref{#1}}}
\newcommand{\sK}[1]{\mbox{(Sect.~\ref{#1})}}

\newcommand{\cpd}{CPD$-20^\circ1123$\xspace}

\title[NLTE spectral analysis of \cpd]{NLTE spectral analysis of the intermediate helium-rich subdwarf B star \cpd\thanks{Based on data products from observations made with ESO Telescopes at the La Silla Paranal Observatory under programme ID 086.D-0714(A).}}

\author[L\@. L\"obling]{L\@. L\"obling$^{1}$\thanks{E-mail: loebling@astro.uni-tuebingen.de}
\\
$^{1}$Institute for Astronomy and Astrophysics, Kepler Center for Astro and Particle Physics,
Eberhard Karls University, \\Sand 1, 72076 T\"ubingen, Germany \\
}

\date{Accepted 2020 June 3. Received 2020 May 29; in original form 2019 August 21.}

\pubyear{2019}

\begin{document}
\label{firstpage}
\pagerange{\pageref{firstpage}--\pageref{lastpage}}
\maketitle
\defcitealias{naslimetal2012}{N12}
\begin{abstract}
Subdwarf B (sdB) stars are core helium-burning stars with stratified atmospheres. 
{Their atmospheres are dominated by hydrogen (H) while the helium (He) and metal abundances are shaped by an interplay of gravitational settling and radiative levitation.}
However, a small fraction of these show spectra dominated by \Ion{He}{1} absorption lines. In between these groups of He-deficient and extreme He-rich sdBs, some are found to have intermediate surface He abundances. These objects are proposed to be young ``normal'' (He-deficient) sdBs for which the dynamical stratification of the atmosphere is still ongoing.
We present an analysis of the optical spectrum of such an intermediate He-rich sdB, namely \cpd, by means of non-local thermodynamic equilibrium (NLTE) stellar atmosphere models. {It has a He-to-H number ratio of $\mathrm{He/H} = 0.13 \pm 0.05$ and its effective temperature of \Teffw{25\,500 \pm 1\,000} together with a surface gravity of $\log\,(g$\,/\,cm/s$^2) = 5.3 \pm 0.3$} places the star close to the high-temperature edge until which it may be justified to use LTE model atmospheres. This work states a test of the T\"ubingen NLTE Model Atmosphere Package for this temperature regime. We present the first application of revised, elaborated model atoms of low ionization stages of light metals usable with this atmosphere code. 
\end{abstract}


%%%%%%%%%%%%%%%%%%%%%%%%%%%%%%%%%%%%%%%%%%%%%%%%%%

%%%%%%%%%%%%%%%%% BODY OF PAPER %%%%%%%%%%%%%%%%%%

\section{Introduction}
\label{sect:intro}

Subdwarf B (sdB) stars are located at the high-temperature end of the horizontal branch (HB) or have evolved off the HB and hence are stars with a central helium (He) burning region. They have thin hydrogen (H) envelopes and in most cases, their atmospheres are He deficient due to gravitational settling. {The abundances of metals are the result of the interaction of gravitational settling and radiative levitation \citep{heber2016}.} However, {a fraction of} 5\,\% \citep{ahmadetal2006} to 13\,\% \citep{greenetal1986} of the total subdwarf population show He-rich atmospheres. Among these He-subdwarf stars (He-sds), a small number {has intermediate} He abundances between 5\,\% and 80\,\% by number \citep{naslimetal2010}. These stars are regarded as transition objects evolving from a He-rich progenitor to a He-poor subdwarf star and, thus, are of particular interest \citep{heber2016}.\\
\cpd \citep[Albus\,1, ][]{gilletal1896,vennes2007} is a blue star and one of the brightest known He-sdB stars. Based on photometric data only, \citet{caballero2007} proposed that it might be a hot white dwarf (WD) similar to the DA-type WD G191$-$B2B or alternatively a hot subdwarf. Using low-resolution optical spectra, \citet{vennes2007} could confirm the subdwarf {classification and found a He enrichment}. Using NLTE model atmospheres, they found \Teffw{19\,800 \pm 400}, \loggw{4.55 \pm 0.10}, and a He abundance of {$\log {\mathrm{He}}/{\mathrm{H}} = 0.15 \pm 0.15$}. \citet[][hereafter \citetalias{naslimetal2012}]{naslimetal2012} had access to high-resolution optical spectra and performed a comprehensive spectral analysis based on LTE model atmospheres. They determined element abundances for elements up to Fe and found \Teffw{23\,500 \pm 500}, \loggw{4.9 \pm 0.1}, and a He abundance of $0.17 \pm 0.05$ based on LTE atmospheres. With a series of radial velocity measurements, they determined an orbital period of the binary to be 2.3\,d and argued that the system is a close binary that underwent common-envelope evolution. Its intermediate He abundance can thus be an indicator {for a} young sdB star in which the diffusion driven stratification of the atmosphere is not yet in equilibrium but still ongoing.\\
For the spectral analysis by means of model atmosphere techniques, a close correspondence between the calculated model and the stellar atmosphere is necessary. Compromises and simplifications need to be taken into consideration due to limitations on computational capacity and available atomic data. While it is highly recommended to use non-local thermodynamic equilibrium (NLTE) model atmospheres for hot white dwarfs and subdwarf O (sdO) stars, it may be justified to use LTE atmospheres for the analysis of sdB stars \citep{napiwotzki1997}. The departure from LTE should be small in this regime due to high densities in the atmosphere. However, \cpd is located at the high-\Teff border of the domain of LTE codes. This implies, that NLTE codes are not frequently used and tested for this regime.\\
This work presents a NLTE analysis of the optical spectrum of \cpd and states a test of the T\"ubingen Model Atmosphere Package \citep[TMAP\footnote{\url{http://astro.uni-tuebingen.de/~TMAP}},][]{werneretal2003a,tmap2012} for this temperature and gravity regime. We describe the observations in \se{sect:obs} {and introduce the model-atmosphere code as well as the revisions and updates that were done on the T\"ubingen Model Atom Database \citep[TMAD,][]{rauchdeetjen2003} in \se{sect:models}. The analysis is described in \se{sect:para} and \se{sect:abund}. The results are discussed in \se{sect:disc} and we end with our conclusions in \se{sect:concl}}.

\section{Observations}
\label{sect:obs}

We use high-resolution, high signal-to-noise observations for this analysis. The spectra were partly processed with a low-pass filter \citep{savitzkygolay1964}. All synthetic spectra shown in this paper are convolved with Gaussians to simulate the instruments' resolving power $R= \lambda\,/\,\Delta\lambda$.\\
Six optical spectra were obtained with the University College London Echelle Spectrograph (UCLES) at the {Anglo-Australian} Telescope (AAT) on 2010-01-14 ($3820\,${\AA}$ < \lambda < 5230\,${\AA}, grating with $31\,\mathrm{lines}\,\mathrm{mm}^{-1}$, $R \approx 45\,000$) with a total exposure time of $9000\,$s. Details about calibration and and post-processing are given by \citetalias{naslimetal2012}.\\
In addition, we used {six} high-resolution spectra taken with the Fiberfed Extended Range Optical Spectrograph (FEROS) at the MPG/ESO 2.2\,m telescope ($3700\,${\AA}$ < \lambda < 9200\,${\AA}, $R \approx 48\,000$) on {2010-10-30 ($700\,$s) and 2010-11-01 ($2 \times 450\,$s, $2 \times 600\,$s, and $900\,$s)}.  {These were retrieved as raw data from the ESO science archive and reduced with the standard ESO MIDAS pipeline which does not perform a background subtraction.}

\section{Model atmospheres and atomic data}
\label{sect:models}

For our spectral analysis, we calculated a grid of stellar-atmosphere models using TMAP. This code calculates chemically homogeneous NLTE models under the assumption of plane-parallel geometry in hydrostatic and radiative equilibrium. Level dissolution (pressure ionization) is considered following \citet{hummermihalas1988} and \citet{hubenyetal1994} for all species. The \Ion{H}{1} line profiles were calculated using the line broadening tables of \citet[][extended tables of 2015, priv\@. comm\@.]{tremblaybergeron2009}. To consider the line broadening of \Ionww{He}{1}{4026, 4338, 4471, 4921}, tables of \citet{bcs1969} and \citet[][only for \Ionw{He}{1}{4471}, for electron densities $10^{13} < n_e\,/\,\mathrm{cm}^3 < 10^{16}$]{bcs1974} were used. For the lines \Ionww{He}{1}{4121, 4438, 4713, 5017, 5048} we used broadening tables of \citet[][]{1974slbp.book.....Griem} and for all other lines in the range $3732\,{\AA} < \lambda < 8997\,{\AA}$ we used the tables of \citet[][]{1997ApJS..108..559Beauchamp}. The line profiles of \Ionww{He}{1}{4144, 4438, 8915} and of all lines outside of the region mentioned above were calculated using an approximate formula for the quadratic Stark effect with parameters of \citet{ds_HeI_1990}. {The lines \Ionww{He}{1}{3926, 4009, 4143} feature in the \citet[][]{1997ApJS..108..559Beauchamp} tables but the modeled line shapes do not agree with the observations. Most likely, our treatment of the physics of these lines is incomplete but since these lines are not used for any parameter determination, they do not affect the results.}
Line broadening tables for some \Ion{Al}{3} lines are available from \citet{ds_alIII_1993} and used for the calculation of the line profiles. Broadening of all other Al lines due to the quadratic Stark effect is calculated using approximate formulae by \citet{cowley1970,cowley1971}. Atomic data for the model calculations {were} obtained from TMAD. This database was enlarged and extended in the course of this analysis \sK{subsect:newatoms}. For the {iron-group elements (IGEs)} (Ca$-$Ni), we used Kurucz's line lists \citep{kurucz1991,kurucz2009} and constructed statistical model atoms using the T\"ubingen Iron-Group Opacity (TIRO) - WWW Interface.

\subsection{New and revised model atoms}
\label{subsect:newatoms}

To reproduce the variety of metal lines found in the optical spectrum of \cpd \citepalias{naslimetal2012}, we constructed new model atoms for aluminum, phosphorus, and chlorine and revised and substantially extended the low ionization stages (\textsc{ii}$-$\textsc{iv}) of the fluorine, sulfur (only for spectrum synthesis), and argon model atoms.\\
The majority of level energies were retrieved from the National Standards and Technology Institute (NIST) Atomic Spectra Database (ASD\footnote{\url{https://physics.nist.gov/PhysRefData/ASD/levels_form.html}}). Oscillator strengths, photoionization crosssections, and additional energy levels were calculated in the framework of the Opacity Project \citep[OP\footnote{\url{http://cdsweb.u-strasbg.fr/topbase/topbase.html}}, ][]{1987Seaton}. Additional sources of data were \citet{eriksson1983} and \citet{beckerbutler1989} for \Ion{N}{2}, and \citet{wieseetal1969}, \citet{fuhrwiese1998}, \citet{bengtson1968}, \citet{foster1962}, \citet{varsavsky1961}, and \citet{maarefetal2012} for Cl\,\textsc{ii}$-$\textsc{iv}. Higher ionization stages of the new model atoms have already been used for the analysis of hot white dwarfs \citep[e.g.,][]{rauch2017}.

\section{Atmospheric parameters}
\label{sect:para}
\begin{figure*}
  \resizebox{\hsize}{!}{\includegraphics{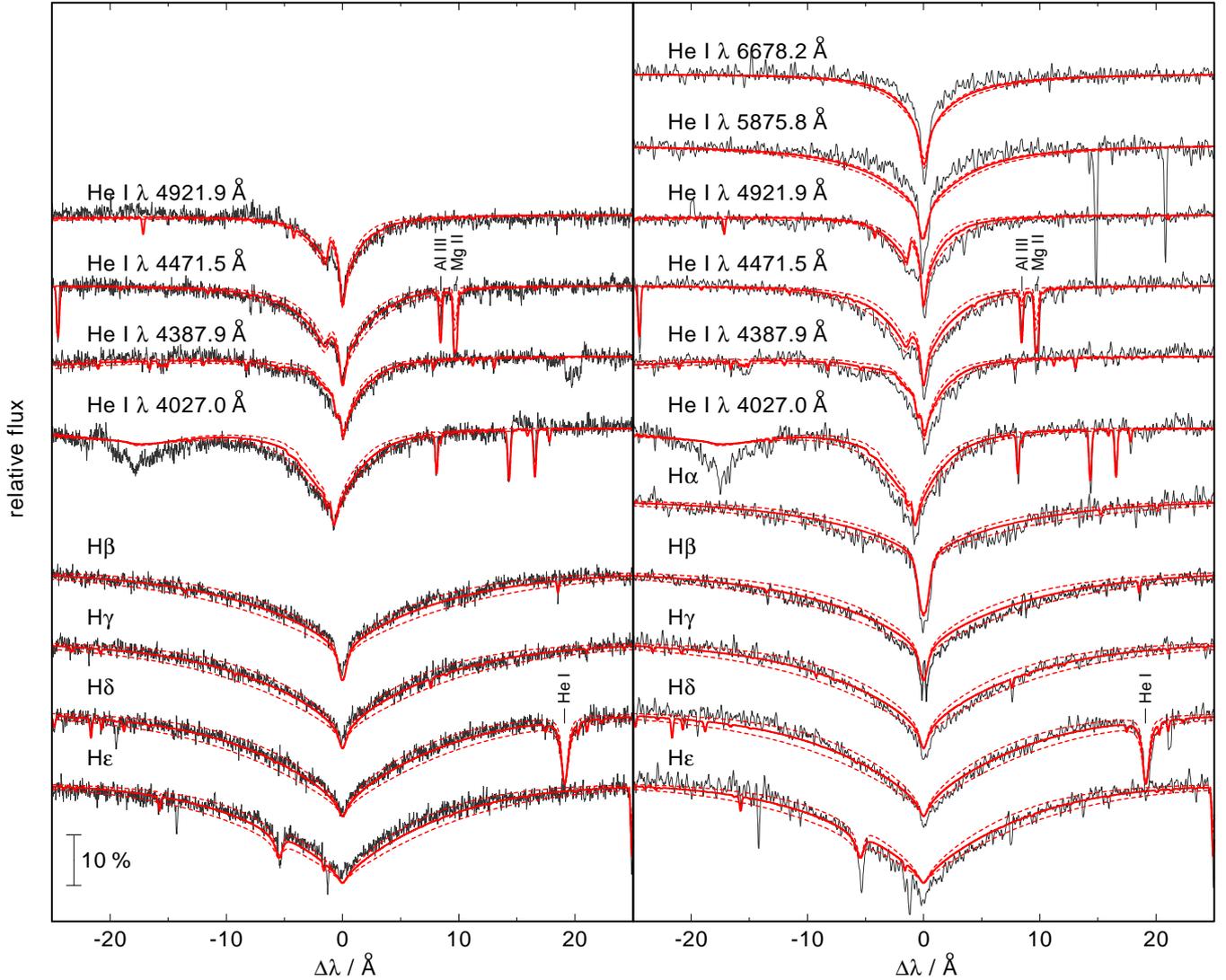}} 
   \caption[]{{Synthetic spectra calculated with \Teffw{25\,500} and \loggw{5.3} (red, solid) and $\Delta \log g = 0.4$ (red, dotted), compared with the UCLES (left) and FEROS (right) observations of \Ionw{He}{1}{4027.0, 4387.9, 4471.5, 4921.9, 5875.8, 6678.2}, H$\alpha$, H$\beta$, H$\gamma$, H$\delta$, and H$\epsilon$.} The vertical bar indicates 10\,\% of the continuum flux.} 
   \label{fig:logg}
\end{figure*}

\begin{figure}
  \resizebox{\hsize}{!}{\includegraphics{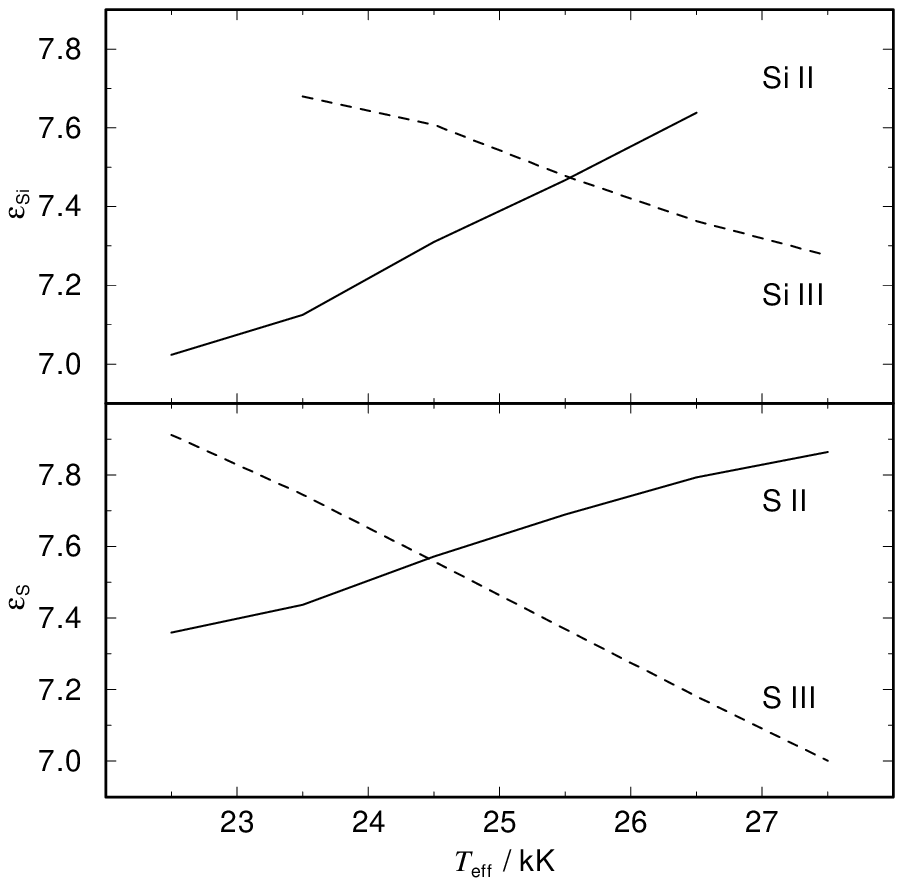}} 
   \caption{{Abundance $\varepsilon_i$ measured based on \Ion{Si}{2} and \Ion{Si}{3} lines (top panel) and on \Ion{S}{2} and \Ion{S}{3} lines (bottom) as a function of temperature for $\logg = 5.3$.}}
   \label{fig:teff}
\end{figure}

The model grid calculated for this analysis spans around the literature values for \Teff, \logg,  {and $\mathrm{He/H}$} of \citetalias{naslimetal2012} ($\Teff = 22\,500\,(1000)\,27\,500$, $\logg = 4.3\,(0.2)\,5.9$, {and $\mathrm{He/H} = 0.1\,(0.1)\,0.3$}) considering the opacities of 22 elements. The statistics of the {model atoms used} are shown in \ta{tab:stat} and the ionization fractions in \ab{fig:ionfrac}. The initial {metal} abundances are taken from \citetalias{naslimetal2012} {while the} abundances for the IGEs {not yet analysed} were scaled to Fe. For F, we adopted the solar value.\\
{Like \citetalias{naslimetal2012}, we used models with a microturbuent velocity of $v_t = 10\,\mathrm{km}\,\mathrm{s}^{-1}$. The atmospheric parameters \Teff, \logg, and the He/H ratio, were determined from a simultaneous fit of H and He lines. The line list of \citetalias{naslimetal2012} has been extended by lines of \citet[][]{schindewolf2018}. The final list is summarized in \ta{tab:hhelines}. The best fit yields \Teffw{25\,500\pm 1000}, \loggw{5.3\pm 0.3}, and $\mathrm{He}/\mathrm{H} = 0.13\pm 0.05$. 
These values differ from the parameters determined by \citetalias{naslimetal2012} (\Teffw{23\,500 \pm 500}, \loggw{4.9 \pm 0.1}, and $\mathrm{He}/\mathrm{H} = 0.17\pm 0.05$). They used a smaller set of Stark-broadenend He and H lines to derive the surface gravity. Figure\,\ref{fig:logg} illustrates that these lines in the UCLES spectrum are not fit equally well by the synthetic spectra with a given \logg. 
By employing also the FEROS observations, we could enlarge the list of broad lines but a simultaneous fit of the data of the two instruments could not be achieved since the observations show significant deviations in the lines that are covered by both. Compared to the UCLES observation, the line cores in the synthetic spectra appear to be too deep. This problem was also seen for the LTE models of \citetalias{naslimetal2012}. Whereas the line cores in the FEROS observation are deeper and the synthetic spectra are too shallow. This may be an effect of different data reduction procedures. While the UCLES observations are background subtracted, the standard FEROS pipeline does no background subtraction. We speculate that background subtraction for the FEROS spectra would even increase the discrepancy. Our determination of the atmospheric parameters is, thus, entirely based on the UCLES spectrum. The FEROS observations are used for cross-check and for abundance determinations in wavelength ranges not covered by UCLES.\\
\citetalias{naslimetal2012} used the ionization equilibria of Si\,\textsc{ii}/\textsc{iii} and S\,\textsc{ii}/\textsc{iii} to determine \Teff and reported that they do not coincide in a single point but that Si\,\textsc{ii}/\textsc{iii} yields a higher temperature than S\,\textsc{ii}/\textsc{iii}. Figure\,\ref{fig:teff} illustrates that this is seen in the same way in our NLTE analysis, although we find higher temperatures. The best \Teff for the Si\,\textsc{ii}/\textsc{iii} lines is in very good agreement with the best fit from the H and He lines, while the S\,\textsc{ii}/\textsc{iii} lines would require a \Teff that is lower by approximately 1000\,K.
}
A former NLTE analysis by \citet{vennes2007} yielded lower values of \Teffw{19\,800 \pm 400} and \loggw{4.55 \pm 0.1} based on a low-resolution optical spectrum. The discrepancy might arise {from} the fact that they used the Balmer and \Ion{He}{1} line-profile fits for the determination of both, \Teff and \logg and the \Ion{H}{1} Stark-broadening tables of \citet{lemke1997} or might be an effect of the borders of the model grid and an inclusion of higher temperatures could have {resulted in parameters similar to this paper}. {Most likely}{, the use of models} containing H and He {in contrast to models that consider} opacities of 22 elements {might have an effect on the final result}. {But still in the present analysis, the Balmer line fits (Figure\,\ref{fig:logg}) indicate missing opacity. The missing flux in the UV indicated by the GALEX observations (Figure\,\ref{fig:ebv}) confirms this explanation.}

\section{Abundances}
\label{sect:abund}

{We adopted \Teffw{25\,500} and \loggw{5.30} for the abundance measurements.} Based on the model grid with 22 elements (statistics given in \ta{tab:stat}), we performed line-formation calculations for each element individually. {In these, the} number of NLTE levels is enlarged and the temperature structure is kept fixed to calculate the occupation numbers for the levels. The abundances were derived by line-profile fits and are affected by typical errors of 0.3\,dex including {the statistical error from the use of several lines and} the error propagation from our uncertainties in \Teff and \logg. For some elements, no line could be identified in the observation. In these cases, we derived upper abundance limits by reducing the abundance until the most prominent computed lines become undetectable within the noise of the spectrum. The resulting abundances are summarized in \ta{tab:finab} and \ab{fig:abundpattern}. The whole UCLES spectrum compared to our final model is shown in \ab{fig:UCLES}.

\begin{figure}
  \resizebox{\hsize}{!}{\includegraphics{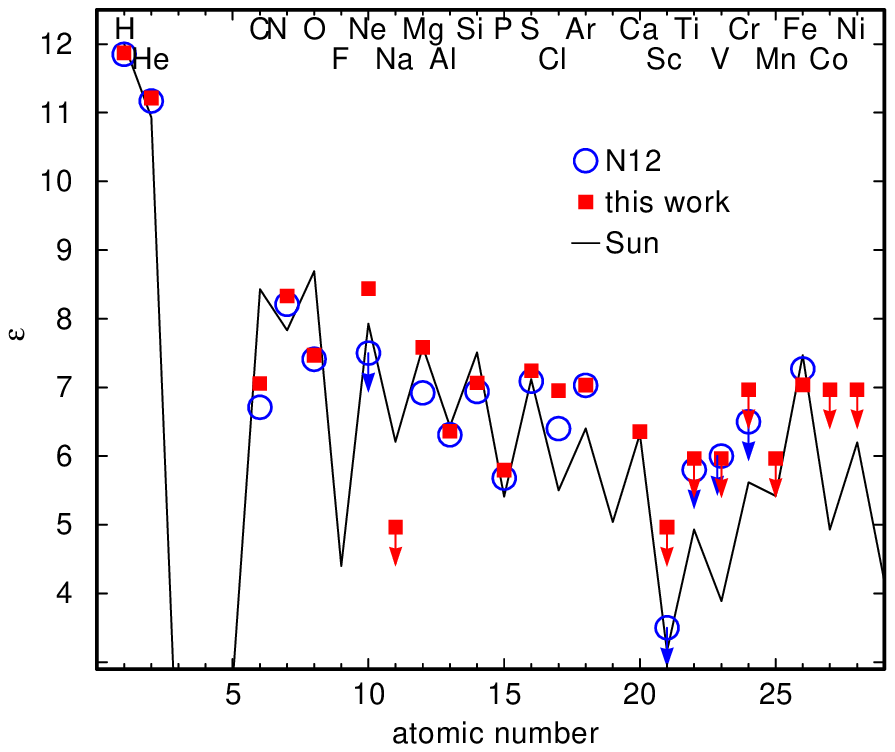}} 
   \caption[]{Photospheric abundances $\epsilon_i$ of \cpd determined from line profile fits compared to the results from \citetalias{naslimetal2012} and solar values from \citet{asplundetal2009,scottetal2015a,scottetal2015b,grevesseetal2015}. Upper limits are indicated with arrows.} 
   \label{fig:abundpattern}
\end{figure}
\begin{figure*}
  \resizebox{\hsize}{!}{\includegraphics{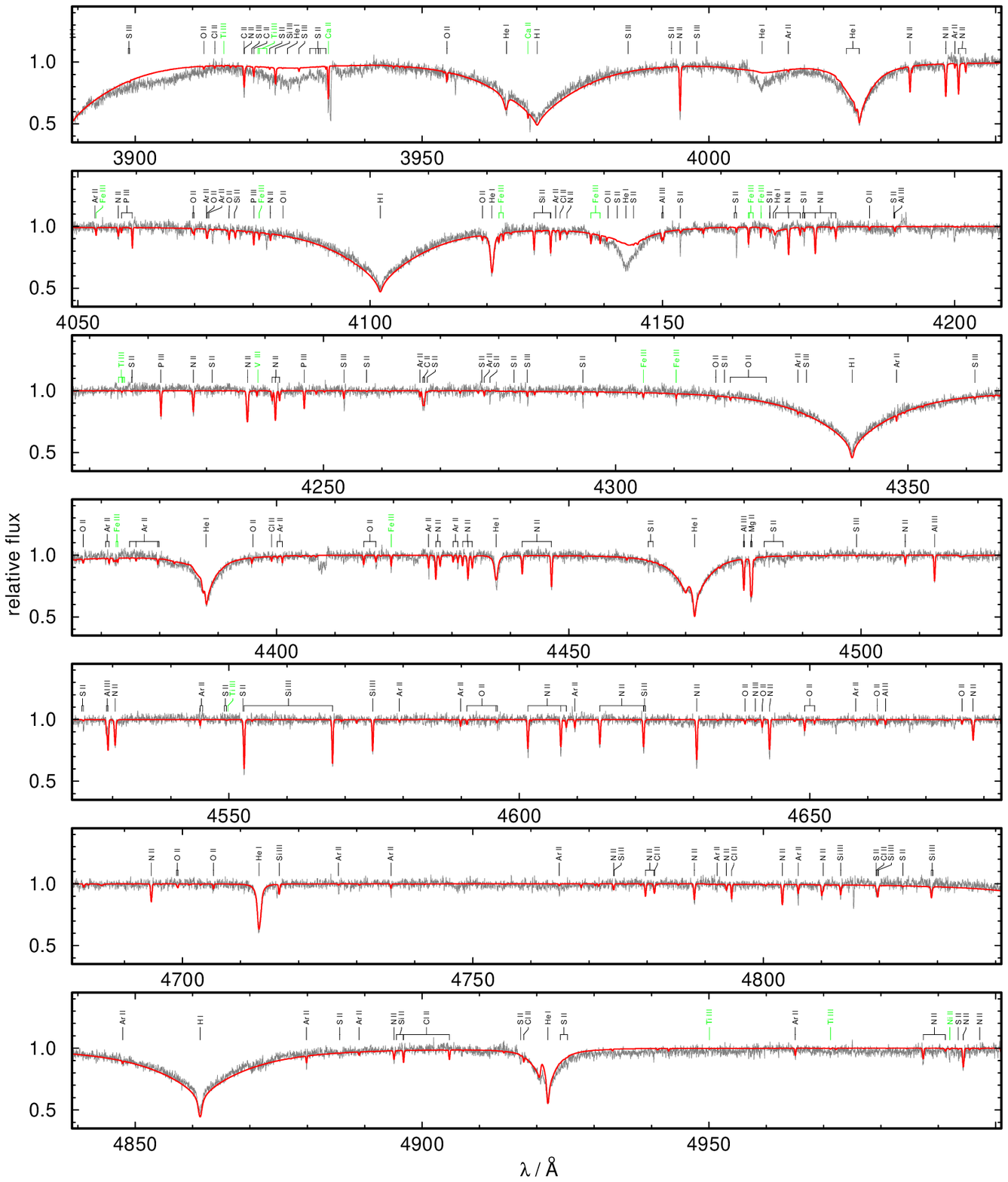}} 
   \caption[]{UCLES spectrum of \cpd compared with the final model with {\Teffw{25\,500} and \loggw{5.30}}. The abundances are given in \ta{tab:finab}. Line-identification marks are included on top, IGEs are indicated in green.} 
   \label{fig:UCLES}
\end{figure*}

\subsection{Light metals}
\label{subsect:lightmetals}

For our abundance measurements, we used the same lines {as} \citetalias{naslimetal2012}. We could not identify any line of F but included it in the atmosphere calculation with solar abundances to take its opacity into account. No lines of Na are present in the observations but an upper limit of $[\mathrm{Na}] \leq -1.26$ (with $[\mathrm{X}] = \log (\mathrm{mass~fraction}/\mathrm{solar~mass~fraction})$) can be derived based on \Ionww{Na}{1}{5148.8, 5153.4} in the FEROS range. The Ne abundance of {$[\mathrm{Ne}] = 0.23$}  is determined based on \Ionw{Ne}{1}{6402.2}. The individual abundance for \Ionww{Al}{3}{4479.9, 4480.0} is omitted from the mean since it yields a significantly lower abundance compared to the result for all other analyzed lines of Al \sA{fig:alplines}. This is contrary to the result of \citetalias{naslimetal2012} who found the largest individual line abundance for these lines. We omitted the line \Ionw{P}{3}{4059.3} from our analysis due to the non-detection of \Ionw{P}{3}{4080.1} which belongs to the same multiplet \sA{fig:alplines}. We include \Ionw{P}{3}{4246.7} which was not considered by \citetalias{naslimetal2012}. For Cl, they report a significantly higher abundance for \Ionw{Cl}{2}{4896.8} compared to \Ionww{Cl}{2}{4794.6, 4810.1} and, thus, omitted the first line. We see overall higher abundances for each line but find the lowest value for \Ionw{Cl}{2}{4896.8} (\ta{tab:cllines}, \ab{fig:cllines}). {This trend of higher abundances in our analysis compared to the results of \citetalias{naslimetal2012} is seen for all light metals with the only exception of Ar.}

\begin{table}
\centering
{
\caption[]{Abundances $\epsilon_i$ for analyzed lines of \Ion{Cl}{2} and \Ion{Fe}{3}.}
\label{tab:cllines}
\begin{tabular}{l@{~~~~~~~}ccc}
\hline
\hline
\noalign{\smallskip}
&& \multicolumn{2}{c}{Abundance} \\
Ion & $\lambda\,/\,${\AA} & This work & \citetalias{naslimetal2012} \\
\noalign{\smallskip}
\hline
\noalign{\smallskip}
\Ion{Cl}{2} & 4794.6 & 7.24  & 6.38  \\
            & 4810.1 & 7.36  & 6.43  \\
            & 4896.8 & 6.94  & 6.96 \\
\noalign{\smallskip}
\Ion{Fe}{3} & 3954.3& 7.34 & \\
            & 4053.1& 6.83 & \\
            & 4081.0& 7.07 & \\
            & 4122.0& 6.80 & \\
            & 4137.8& 7.12 & \\
            & 4139.4& 6.80 & \\
            & 4164.7& 6.97  & 7.13 \\
            & 4166.8& 7.17  & 7.38 \\
            & 4238.6& 7.34 & \\
            & 4304.8& 7.64 & \\
            & 4310.4& 7.81 & \\
            & 4395.8& 7.17  & 7.16 \\
            & 4419.6& 7.61  & 7.30 \\
            & 4431.0& 7.73 & \\
\hline
\end{tabular}
\newline
%\raggedright{
\textbf{Notes. }
Abundances in the form {$\epsilon_i = \log \left( n_i / n_\mathrm{H} \right) + 12$, \newline with the number densities $n_i$ and $n_\mathrm{H}$ of elements $i$ and H.}
%}
}
\end{table}

\begin{figure}
  \resizebox{\hsize}{!}{\includegraphics{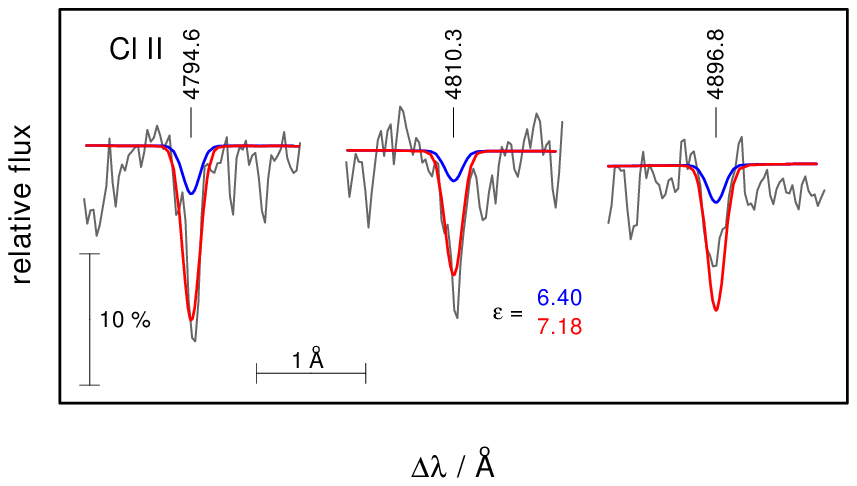}} 
   \caption[]{Synthetic spectra calculated with \loggw{5.30}, $\Teff = 25\,500$, and Cl abundances of $\log \varepsilon = 6.40$ (blue) and 7.18 (red), compared with the UCLES observations of lines of \Ion{Cl}{2}.} 
   \label{fig:cllines}
\end{figure}

\subsection{Iron-group elements}
\label{subsect:ige}

We determine an abundance of {$[\mathrm{Ca}] = 0.67$}  based on \Ionww{Ca}{2}{3933.7, 3968.5}. The Fe abundance of {$[\mathrm{Fe}] = -0.24$} is determined by analyzing 14 lines of \Ion{Fe}{3} \sT{tab:cllines}. This element is the only metal, for which we found a lower abundance compared to the values found in the LTE analysis of \citetalias{naslimetal2012}. This can already be seen when comparing the abundances for the four \Ion{Fe}{3} lines considered by them (\ta{tab:cllines}, \ab{fig:felines}). Again, we find, that there is not only an offset between the different {analyses} but that the discrepancies are different from line to line. 
For all other IGE, we could not identify any line but {derived} upper abundance limits. The values for Ti and V are very close to the ones found by \citetalias{naslimetal2012} whereas our results for Sc and Cr are higher than {theirs}. Mn, Co, and Ni were not analyzed before and we use the computed lines \Ionw{Mn}{3}{4292.9}, \Ionww{Co}{3}{3932.9, 3955.1}, and \Ionww{Ni}{3}{4362.8, 4363.7, 4365.4} to determine upper abundance limits.

\section{Discussion}
\label{sect:disc}

\subsection{{Departures from LTE}}
\label{subsect:nlte}

\begin{figure}
  \resizebox{\hsize}{!}{\includegraphics{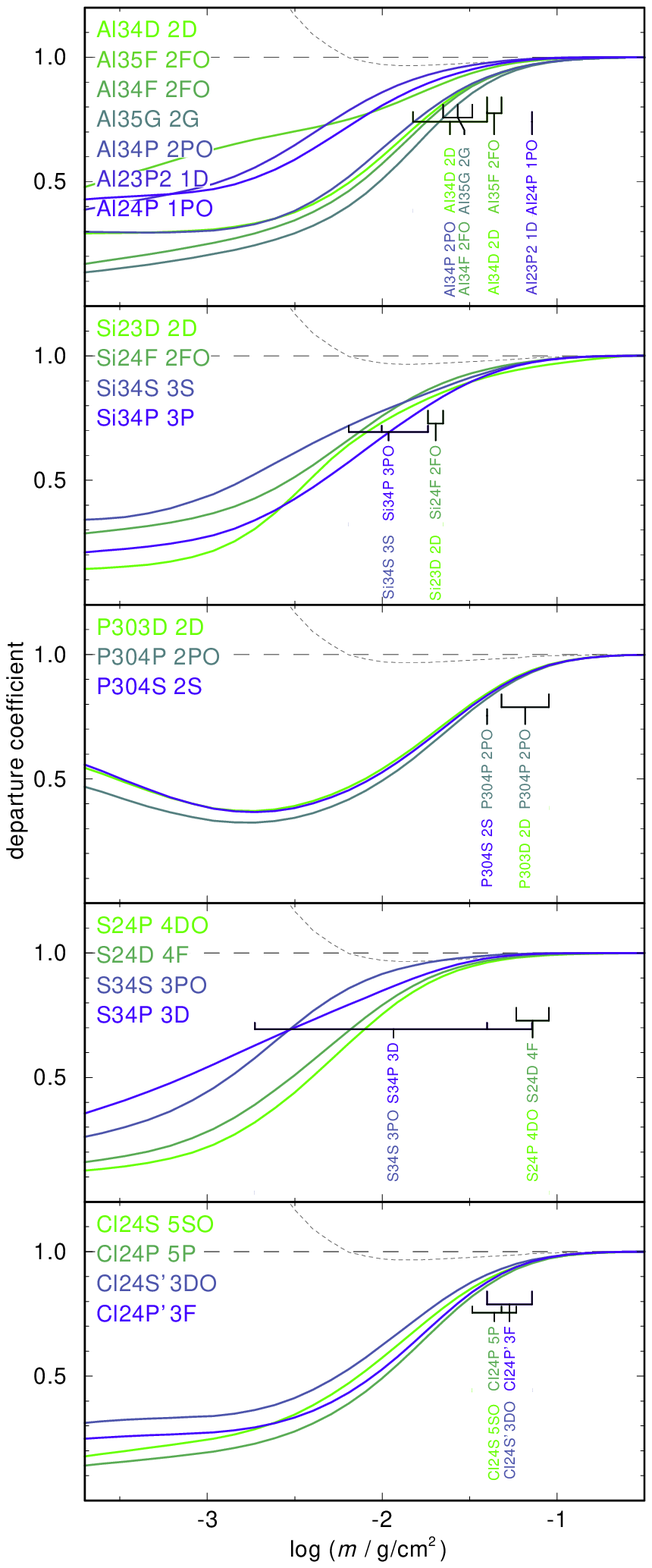}} 
   \caption[]{Departure coefficients of the \Ion{Al}{2}, \Ion{Al}{3}, \Ion{Si}{2}, \Ion{Si}{3}, \Ion{P}{3}, \Ion{S}{2}, \Ion{S}{3}, and \Ion{Cl}{2} levels corresponding to the lines in Figs.\,\ref{fig:teff}, \ref{fig:alplines}, and \ref{fig:cllines}. The formation depth of the line cores is indicated. The departure coefficient of the lowest H level is included (gray, dashed).} 
   \label{fig:sisalpcldep}
\end{figure}

\begin{figure}
  \resizebox{\hsize}{!}{\includegraphics{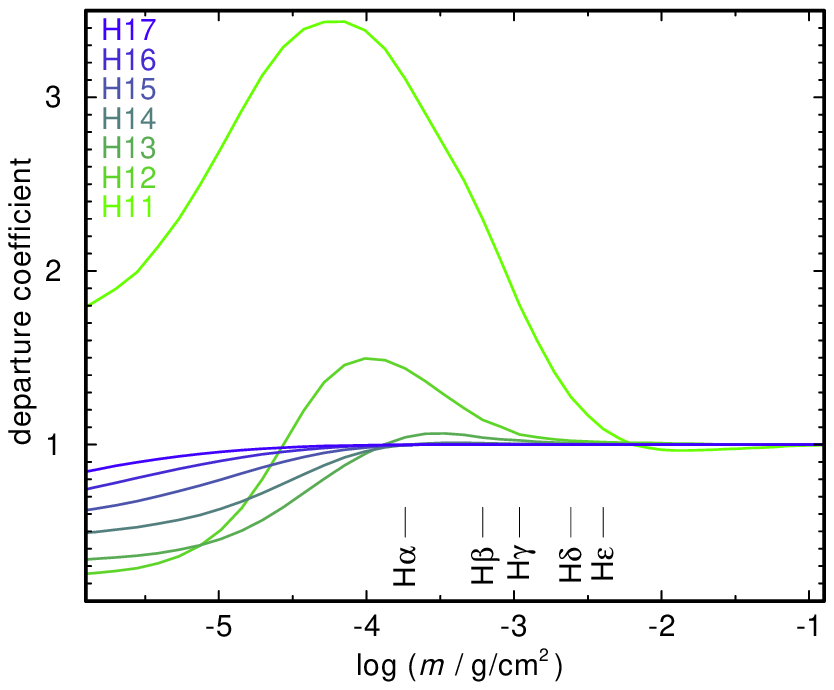}} 
   \caption[]{Departure coefficients of the H levels and formation depth of the cores of the Balmer lines.} 
   \label{fig:hdep}
\end{figure}

{The aim of this work is to enlarge and update TMAD for low ionization stages of light metals and apply the new model atoms in an analysis of an object at the border of the regimes of NLTE and LTE codes, that has already been analyzed comprehensively by means of LTE. The departure coefficients (ratio of NLTE and LTE occupation numbers) are a measure for the importance of NLTE effects. 
Of special interest are the departure coefficients of the new model atoms. Figure\,\ref{fig:sisalpcldep} shows the departure coefficients for the levels of \Ion{Al}{2}, \Ion{Al}{3}, \Ion{P}{3}, and \Ion{Cl}{2} corresponding to the lines in Fig.\,\ref{fig:alplines}, and \ref{fig:cllines}. In the line-forming region, all these coefficients are below unity. 
The speculation that these lower NLTE occupation numbers compared to LTE are responsible for the higher light metal abundances could be ruled out by a LTE model produced with TMAP. This can be achieved in TMAP by artificially increasing the collisional rates in the calculation and thus forcing the departure coefficients to unity. \ab{fig:tstructure} shows the temperature structure of the two different models. Although the ratio $T_\mathrm{NLTE}/T_\mathrm{LTE}$ is close to unity, a slightly lower $T_\mathrm{NLTE}$ is visible in the inner part of the line-forming region while the outermost part shows a higher $T_\mathrm{NLTE}$. The difference in emergent flux between the TMAP NLTE and LTE model does not exceed 2\%, which can by far not explain the abundance discrepancies between the analysis of \citetalias{naslimetal2012} and this work. It needs to be questioned whether the differences arise from different codes or are an effect of the elemental composition of the {model atmosphere grid used}.\\
Figure\,\ref{fig:hdep} shows the departure coefficients of the levels of H. While the departure coefficient of the lowest level is below unity in the deepest part of the line-forming region, it reaches up to three in the outer part. This coefficient determines the relation between the local temperatures of a NLTE and LTE model. If it is below unity, it requires $T_\mathrm{NLTE} < T_\mathrm{LTE}$ \citep{kudritzki1979}.
In the formation depth of the Balmer lines, the departure coefficients of the H levels deviate from each other \sA{fig:hdep}. This might be a reason for discrepancies in the Balmer-line cores that were reported in LTE analyses \citep{napiwotzki1997}.\\
}
As described in \se{sect:para} and shown in \ab{fig:teff}, we see a discrepancy in the temperature determined from Si\,\textsc{ii}/\textsc{iii} and S\,\textsc{ii}/\textsc{iii} that was also reported by \citetalias{naslimetal2012}. 
{To evaluate the importance of NLTE effects for the lines of Si and S and thus on the temperature determination, we investigated also the departure coefficients for the levels corresponding to the lines used for the abundance determinations \sA{fig:teff}. The departure coefficients are below unity in the line-forming region \sA{fig:sisalpcldep}, However, when comparing the TMAD LTE and NLTE models, the line profiles coincide within the thickness of a line. This result validates the well established practice to use LTE codes for the analysis of sdB stars.}

\begin{figure}
  \resizebox{\hsize}{!}{\includegraphics{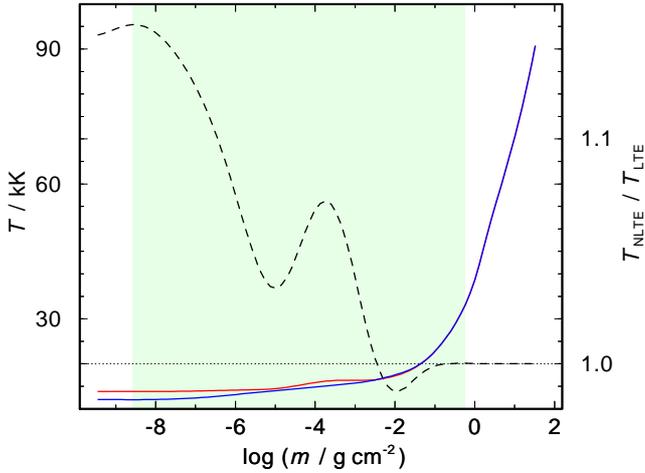}} 
   \caption[]{{Temperature structure of a NLTE (red) and a LTE model (blue). Both are calculated with \Teffw{24\,500} and \loggw{4.9}. $T_\mathrm{NLTE}/T_\mathrm{LTE}$ is indicated as dashed line. The line-forming region is indicated in green.}} 
   \label{fig:tstructure}
\end{figure}

\subsection{Mass, spectral type, and distance}
\label{subsect:typedist}
We determine the mass of \cpd by comparing its position in the \Teff-\logg diagram with evolutionary tracks for sdB stars from \citet[][$Z = Z_\odot$, $Y=Y_\odot$]{dormanetal1993} \sA{fig:tefflogg} and find {$M = 0.475 \pm 0.015\,M_\odot$ which is close to the assumed typical sdB mass of $M = 0.47\,M_\odot$ from \citepalias{naslimetal2012}. The error includes the propagation of uncertainties in \Teff and \logg as well as in the abundances, that we estimated from tracks for different compositions. However, we want to note that this mass determination has to be treated with caution, since, as discussed in Section~\ref{sect:intro}, the intermediate He abundance might indicate a pre-EHB nature of the object. Different pre-EHB evolutionary tracks for the hot-flasher scenario exist \citep[e.g., ][]{2003ApJ...582L..43Cassisi,2008A&A...491..253M3B, 2018A&A...614A.136Battich} but the mass determination from these is difficult since they differ depending on the evolutionary phase when the flash occurs, i.e., whether it is a early or late hot flasher with deep or shallow mixing. Furthermore, these tracks might be inappropriate as well due to the fact that \cpd is in a short period binary with a low-mass main-sequence companion or a white dwarf \citepalias[WD, ][]{naslimetal2012}, which suggests that the sdB star formed via a binary channel and lost its envelope in a mass-transfer event rather than in a hot-flasher scenario.}
From the position in the color-color {diagram} and a spectral energy distribution (SED, \ab{fig:ebv}) {similar to G191$-$B2B}, \citet{caballero2007} classified \cpd as an early DA-type WD. {Both stars'} 2MASS $K_s$ magnitudes are identical but G191$-$B2B is brighter in the B-band. Assuming the same spectral type, they concluded, that \cpd should be colder and, thus, closer to the Sun compared to G191$-$B2B. We find, that our best model for \cpd resembles the observed magnitudes better than a model SED for G191$-$B2B \citep[\Teffw{60\,000}, \loggw{7.6}, obtained from TheoSSA\footnote{\url{http://dc.g-vo.org/theossa}}, \ab{fig:ebv},][]{rauchetal2013}. Including also the GALEX FUV {and NUV} magnitude \citep[][]{bianchietal2017}, we can furthermore {confirm, that the observations are affected by interstellar reddening of \ebvw{0.0684} \citep{2011ApJ...737..103SchlaflyFinkbeiner}}.
With the knowledge of \logg from spectral analyses, \citet{vennes2007} revised the spectral type and classified \cpd as an sdB. Following the same arguments as above, we can now conclude from the identical $K_s$ magnitude, that \cpd should be further away from the Sun compared to G191$-$B2B. This is confirmed by the distances from Gaia parallaxes \citep{baillerjonesetal2018}, namely $329\pm 7$\,pc for \cpd and $52.85\pm 0.19$ for G191$-$B2B. 
{The precise Gaia distance is an excellent test for the atmospheric parameters. With the values from Table\,\ref{tab:finab}, we get a spectroscopic distance of $425^{+128}_{-177}$\,pc which is still in agreement with the Gaia value. The surface gravity determined in previous studies is too low since \loggw{4.9 \pm 0.1} \citepalias{naslimetal2012} would yield a distance of $678^{+87}_{-102}$\,pc.\\
Using the Gaia distance, and our atmospheric parameters, we can test the mass determination from evolutionary tracks. We find $M = 0.24^{+0.24}_{-0.12}$ which is too low for a typical sdB star but just in agreement with the other method. To increase this value to 0.475, a higher surface gravity of \loggw{5.6} would be needed, {and is within the observational error}.
}

\begin{figure}
  \resizebox{\hsize}{!}{\includegraphics{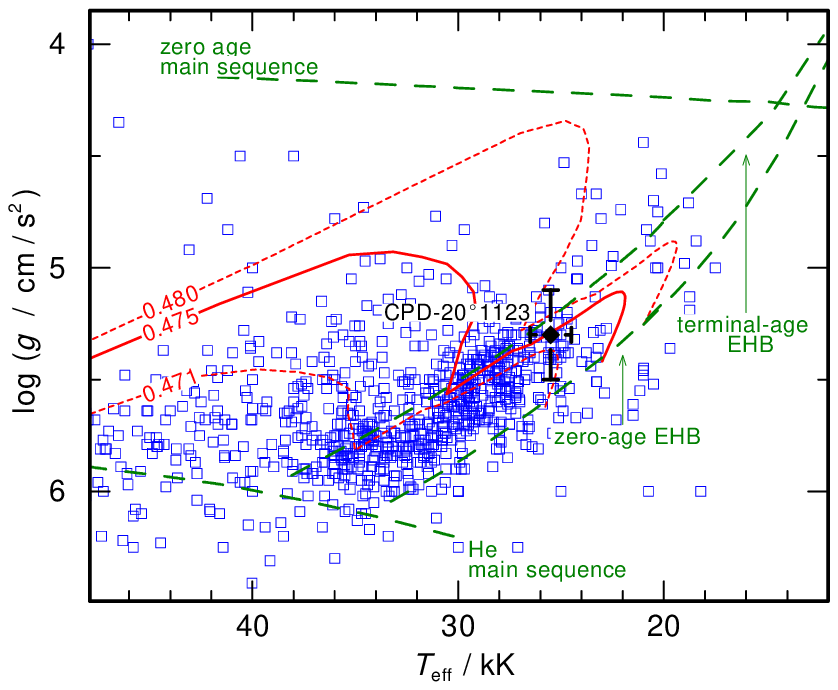}} 
   \caption[]{Location of \cpd (black, with error bars) in the \Teff-\logg diagram compared with sd(O)B-type stars close to the extended horizontal branch (EHB) from \citet[][blue squares]{geiertela2017}. Post-EHB evolutionary tracks from \citet[][red, $Z = Z_\odot$, $Y=Y_\odot$]{dormanetal1993} are included and labeled with stellar mass.} 
   \label{fig:tefflogg}
\end{figure}

\section{Conclusions}
\label{sect:concl}

In the course of this work, we extended our model-atom database TMAD by constructing new model atoms for Al, P, and Cl. The low ionization stages (\textsc{ii}-\textsc{iv}) of F, S, and Ar were substantially enlarged. The analysis of the optical spectrum of the intermediate He-rich sdB star \cpd constitutes a test of NLTE stellar atmosphere code TMAP for the low-\Teff regime. 
This star is located close the high-temperature edge until which it may be justified to use LTE model atmospheres for the spectral analysis. However, NLTE effects are present in every star and become visible, at least, in the analysis of high-resolution spectra and in high-energy observations.
We found that {the departure coefficients differ from unity in the line-forming region. This effect is however not strong enough to cause visible differences in the line profiles comparing NLTE and LTE atmospheres.} 
{We determine \Teffw{25\,500 \pm 1000} which is 2000\,K higher than the value found by \citetalias{naslimetal2012} in their LTE analysis. The surface gravity of \loggw{5.3 \pm 0.3} is also higher by 0.4\,dex which is favored by the Gaia distance measurement. Our abundances for the light metals are higher up to a factor of 7 compared to those of \citetalias{naslimetal2012}. Fe is the only element, for which we determine a slightly lower abundance.
Our comparison of a NLTE and LTE model calculated with TMAP for the same atmospheric parameters and abundances ruled out that these discrepancies are NLTE effects, since the synthetic spectra for these two models differ only marginally.
This work proves that with the presented extension of the atomic database for low ionization stages, NLTE and LTE codes are equally well suited for the analysis of sdB stars. 
}

\section*{Acknowledgements}
The author thanks Thomas Rauch for the idea and initiation of this project and the helpful discussions and comments and Klaus Werner for the help with the draft of this publication. I thank Michael Kn\"orzer and Stefanie Klein for the preparation of some of the new TMAD model atoms.
The author thanks the anonymous referee for the very constructive comments and corrections that improved the quality of this work. I thank Simon Jeffery for the discussions about the object and this work and for the very helpful comments and corrections during the review process.  
We thank Naslim Neelamkodan and Simon Jeffery for putting the reduced UCLES spectra at our disposal. We thank John Pritchard from the ESO User Support Department for the explanations about FEROS and the data reduction process. 
This work was supported by the German Research Foundation (DFG, grant WE\,1312/49-1).
We were supported by the High Performance and Cloud Computing Group 
at the Zentrum f\"ur Datenverarbeitung of the University of T\"ubingen, the state of
Baden-W\"urttemberg through bwHPC, and the DFG (grant INST\,37/935-1\,FUGG).
The GAVO project had been supported by the Federal Ministry of Education and Research (BMBF) 
at T\"ubingen (05\,AC\,6\,VTB, 05\,AC\,11\,VTB).
The TIRO (\url{http://astro-uni-tuebingen.de/~TIRO}),
TMAD (\url{http://astro-uni-tuebingen.de/~TMAD}), and TheoSSA (\url{http://dc.g-vo.org/theossa}) services used for this paper 
were constructed as part of the activities of the German Astrophysical Virtual Observatory.
This research has made use of 
NASA's Astrophysics Data System and
the SIMBAD database, operated at CDS, Strasbourg, France.
This work has made use of data from the European Space Agency (ESA) mission
{\it Gaia} (\url{https://www.cosmos.esa.int/gaia}), processed by the {\it Gaia}
Data Processing and Analysis Consortium (DPAC,
\url{https://www.cosmos.esa.int/web/gaia/dpac/consortium}). Funding for the DPAC
has been provided by national institutions, in particular the institutions
participating in the {\it Gaia} Multilateral Agreement.

%%%%%%%%%%%%%%%%%%%%%%%%%%%%%%%%%%%%%%%%%%%%%%%%%%



\bibliographystyle{mnras}
\bibliography{cpd} % if your bibtex file is called example.bib

%%%%%%%%%%%%%%%%%%%%%%%%%%%%%%%%%%%%%%%%%%%%%%%%%%

%%%%%%%%%%%%%%%%% APPENDICES %%%%%%%%%%%%%%%%%%%%%

\clearpage
\appendix

\section{Additional figures and tables.}
\label{app:additional}
\begin{table*}\centering
\caption{Statistics of the H -- Ar$^{a}$ and Ca - Ni$^{b}$ model atoms used in our model-atmosphere calculations.}
\label{tab:stat}
\setlength{\tabcolsep}{.3em}
\begin{tabular}{rlrrrp{10mm}rlrrr}
\hline
\hline
\multicolumn{2}{l}{}    & \multicolumn{2}{c}{Levels} & & &\multicolumn{2}{l}{}& \multicolumn{1}{l}{Super}                   & \multicolumn{1}{c}{Super} & Individual\\
\cline{3-4}
\multicolumn{2}{l}{}    &      &        &          && \multicolumn{2}{l}{}    &                         &        &       \vspace{-5.5mm}\\
\multicolumn{2}{l}{\hbox{}\hspace{4mm}Ion} &      &        &  ~Lines  && 
\multicolumn{2}{l}{\hbox{}\hspace{4mm}Ion} &        &        &       \vspace{-1.5mm}\\
\multicolumn{2}{l}{}    & NLTE & ~~~LTE &          && \multicolumn{2}{l}{}    & \multicolumn{1}{c}{levels$^{c}$} & \multicolumn{1}{c}{lines}  & \multicolumn{1}{c}{lines}                \\
\hline
\noalign{\smallskip}
H  & \Ion{}{1} &         10   &    22   &    45   &   &     Ca & \Ion{}{2} &         7   &   26  &      2\,612\\
   & \Ion{}{2} &          1   &   $-$   &   $-$   &   &        & \Ion{}{3} &         7   &   26  &     40\,664\\
He & \Ion{}{1} &        103   &     0   &   504   &   &        & \Ion{}{4} &         1   &    0  &           0\\
   & \Ion{}{2} &         16   &    16   &   120   &   &     Sc & \Ion{}{2} &         7   &   26  &     77\,014\\
   & \Ion{}{3} &          1   &   $-$   &   $-$   &   &        & \Ion{}{3} &         7   &   26  &      1\,299\\
C  & \Ion{}{2} &         10   &    36   &    14   &   &        & \Ion{}{4} &         1   &    0  &           0\\
   & \Ion{}{3} &         16   &    89   &    30   &   &     Ti & \Ion{}{2} &         7   &   27  &    312\,054\\
   & \Ion{}{4} &          1   &     0   &     0   &   &        & \Ion{}{3} &         7   &   25  &     46\,707\\
N  & \Ion{}{1} &         10   &    17   &    11   &   &        & \Ion{}{4} &         7   &   24  &      2\,226\\
   & \Ion{}{2} &         16   &   232   &    24   &   &        & \Ion{}{5} &         1   &    0  &           0\\
   & \Ion{}{3} &         16   &    50   &    34   &   &     Cr & \Ion{}{2} &         7   &   27  &    728\,080\\
   & \Ion{}{4} &          1   &     0   &     0   &   &        & \Ion{}{3} &         7   &   27  & 1\,421\,382\\
O  & \Ion{}{1} &         10   &    86   &     7   &   &        & \Ion{}{4} &         7   &   23  &    234\,170\\
   & \Ion{}{2} &         16   &    31   &    26   &   &        & \Ion{}{5} &         1   &    0  &           0\\
   & \Ion{}{3} &          1   &     0   &     0   &   &     Mn & \Ion{}{2} &         7   &   27  &    136\,814\\
F  & \Ion{}{2} &         10   &    98   &     8   &   &        & \Ion{}{3} &         7   &   27  & 1\,668\,146\\
   & \Ion{}{3} &          1   &     0   &     0   &   &        & \Ion{}{4} &         7   &   25  &    719\,387\\
Ne & \Ion{}{1} &          6   &    32   &     3   &   &        & \Ion{}{5} &         1   &    0  &           0\\
   & \Ion{}{2} &         15   &    19   &    21   &   &     Fe & \Ion{}{2} &         7   &   27  &    531\,460\\
   & \Ion{}{3} &          1   &     0   &     0   &   &        & \Ion{}{3} &         7   &   25  &    537\,689\\
Mg & \Ion{}{2} &         14   &    16   &    34   &   &        & \Ion{}{4} &         7   &   25  & 3\,102\,371\\
   & \Ion{}{3} &          1   &     0   &     0   &   &        & \Ion{}{5} &         1   &    0  &           0\\
Al & \Ion{}{2} &         10   &   118   &    12   &   &     Co & \Ion{}{2} &         7   &   27  &    593\,140\\
   & \Ion{}{3} &         12   &    28   &    25   &   &        & \Ion{}{3} &         7   &   27  & 1\,325\,205\\
   & \Ion{}{4} &          1   &     0   &     0   &   &        & \Ion{}{4} &         7   &   27  &    552\,916\\
Si & \Ion{}{2} &         13   &    45   &    28   &   &        & \Ion{}{5} &         1   &    0  &           0\\
   & \Ion{}{3} &         16   &    94   &    25   &   &     Ni & \Ion{}{2} &         7   &   27  &    322\,269\\
   & \Ion{}{4} &          1   &     0   &     0   &   &        & \Ion{}{3} &         7   &   22  & 1\,033\,920\\
P  & \Ion{}{2} &          8   &    38   &     2   &   &        & \Ion{}{4} &         7   &   25  & 2\,512\,561\\
   & \Ion{}{3} &         15   &    39   &     4   &   &        & \Ion{}{5} &         1   &    0  &           0\\
   & \Ion{}{4} &          1   &     0   &     0    &  &         &           &            &       &            \\ 
S  & \Ion{}{2} &         16   &   220   &    18    &  & 	     & 		 & 	      &       & 	 \\
   & \Ion{}{3} &         15   &   216   &    12    &  & 	     & 		 & 	      &       & 	 \\
   & \Ion{}{4} &          1   &     0   &     0    &  & 	     & 		 & 	      &       & 	 \\
Cl & \Ion{}{2} &         11   &    75   &     5    &  & 	     & 		 & 	      &       & 	 \\
   & \Ion{}{3} &         16   &    18   &    15    &  & 	     & 		 & 	      &       & 	 \\
   & \Ion{}{4} &          1   &     0   &     0    &  & 	     & 		 & 	      &       & 	 \\
Ar & \Ion{}{2} &         10   &   238   &     4    &  & 	     & 		 & 	      &       & 	 \\
   & \Ion{}{3} &         15   &   268   &     6    &  & 	     & 		 & 	      &       & 	 \\
   & \Ion{}{4} &          1   &     0   &     0    &  &         &           &            &       &          \\     
\cline{1-11}
\noalign{\smallskip}
total &        &  446 & 2141 &  1037 &&                     &           &  155 &   568 &  15\,902\,086 \\
\hline
\end{tabular}

\textbf{Notes.} $^{(a)}${classical model atoms},
$^{(b)}${model atoms constructed using a statistical approach \citep{rauchdeetjen2003}},\\
$^{(c)}${treated as NLTE levels}.
\end{table*}
\begin{table}
\centering
{
\caption[]{H and He lines used for the determination of atmospheric parameters.}
\label{tab:hhelines}
\begin{tabular}{lr}
\hline
\hline
\noalign{\smallskip}
Ion & $\lambda\,/\,${\AA}\\
\noalign{\smallskip}
\hline
\noalign{\smallskip}
\multicolumn{2}{l}{UCLES and FEROS range}\\
\Ion{H}{1}  & 3970.08  \\
\Ion{He}{1}  & 4026.98 \\
\Ion{H}{1}  & 4101.71 \\
\Ion{H}{1}  & 4340.46 \\
\Ion{He}{1}  & 4387.93 \\
\Ion{He}{1}  & 4437.55 \\
\Ion{He}{1}  & 4471.50 \\
%\Ion{He}{2}  & 4685.70 \\
%\Ion{He}{1}  & 4713.17 \\
\Ion{H}{1}  & 4861.32 \\
\Ion{He}{1}  & 4921.93 \\
\noalign{\smallskip}
\multicolumn{2}{l}{only FEROS range}\\
\Ion{He}{1}  & 5015.68 \\
\Ion{He}{1}  & 5047.74 \\
\Ion{He}{1}  & 5875.75 \\
\Ion{H}{1}  & 6562.79 \\
\Ion{He}{1}  & 6678.15 \\
\hline
\end{tabular}
}
\end{table}

\begin{table*}
\setlength{\tabcolsep}{0.80em}
\caption[]{Parameters of \cpd.}
{
\centering
\label{tab:finab}
\begin{tabular}{rrrrrr}
\hline
\hline
\noalign{\smallskip}
$T_\mathrm{eff}\,/\,$K     & \multicolumn{5}{l}{$25\,500 \pm 1\,000$}\\
\noalign{\smallskip}	    
$\log\,(g$\,/\,cm/s$^2$)  &  \multicolumn{5}{l}{$5.30 \pm 0.3 $}\\
\noalign{\smallskip}	    
$d$\,/\,pc                &  \multicolumn{5}{l}{$329\pm 7\,^{(a)}$}\\
\noalign{\smallskip}	     
$M\,/\,M_\odot$            & \multicolumn{5}{l}{$0.475 \pm 0.015\,^{(b)}$}  \\
\noalign{\smallskip}	    
$\ebv$            & \multicolumn{5}{l}{$0.0684 \pm 0.0009\,^{(c)}$}  \\
\noalign{\smallskip}	    
%$\log\ ( L\,/\,L_\odot )$  & \multicolumn{5}{l}{$1.73 \pm 0.15$}  \\      
%\noalign{\smallskip}	    
GALEX FUV        &  \multicolumn{5}{l}{$11.9094 \pm 0.0045$\,$^{(d)}$}\\
GALEX NUV        &  \multicolumn{5}{l}{$12.4470 \pm 0.0038$\,$^{(d)}$}\\
$B$            &  \multicolumn{5}{l}{$11.907 \pm 0.06$\,$^{(e)}$}\\
$V$            &  \multicolumn{5}{l}{$12.173 \pm 0.12$\,$^{(f)}$}\\
$r$              &  \multicolumn{5}{l}{$12.336 \pm 0.01$\,$^{(e)}$}\\
$i$              &  \multicolumn{5}{l}{$12.638 \pm 0.06$\,$^{(e)}$}\\
$J$              &  \multicolumn{5}{l}{$12.56 \pm 0.02$\,$^{(g)}$}\\
$H$              &  \multicolumn{5}{l}{$12.66 \pm 0.03$\,$^{(g)}$}\\
$K_s$            &  \multicolumn{5}{l}{$12.76 \pm 0.03$\,$^{(g)}$}\\
\hline
\noalign{\smallskip}
Abundances & [X] & Mass fraction & Number fraction & $\varepsilon$ & [X/Fe] \\
\noalign{\smallskip}
\hline
H    &   $-0.06$   &   $6.37\times 10^{-1}$   &   $8.77\times 10^{-1}$   &   $11.95$   &   $ 0.21$\\
He   &   $ 0.15$   &   $3.52\times 10^{-1}$   &   $1.22\times 10^{-1}$   &   $11.09$   &   $ 0.42$\\
C    &   $-0.88$   &   $3.12\times 10^{-4}$   &   $3.61\times 10^{-5}$   &   $ 7.57$   &   $-0.61$\\
N    &   $ 0.71$   &   $3.52\times 10^{-3}$   &   $3.49\times 10^{-4}$   &   $ 8.55$   &   $ 0.98$\\
O    &   $-1.16$   &   $3.94\times 10^{-4}$   &   $3.42\times 10^{-5}$   &   $ 7.54$   &   $-0.89$\\
Ne   &   $ 0.23$   &   $2.14\times 10^{-3}$   &   $1.47\times 10^{-4}$   &   $ 8.18$   &   $ 0.50$\\
Na   &$\le-1.26$   &$\le1.50\times 10^{-6}$   &$\le1.00\times 10^{-7}$   &$\le 4.97$   &$\le-0.98$\\
Mg   &   $ 0.18$   &   $1.04\times 10^{-3}$   &   $5.93\times 10^{-5}$   &   $ 7.78$   &   $ 0.44$\\
Al   &   $ 0.47$   &   $1.55\times 10^{-4}$   &   $7.99\times 10^{-6}$   &   $ 6.91$   &   $ 0.72$\\
Si   &   $-0.05$   &   $5.87\times 10^{-4}$   &   $2.90\times 10^{-5}$   &   $ 7.47$   &   $ 0.22$\\
P    &   $ 0.80$   &   $3.64\times 10^{-5}$   &   $1.63\times 10^{-6}$   &   $ 6.22$   &   $ 1.07$\\
S    &   $ 0.43$   &   $8.34\times 10^{-4}$   &   $3.61\times 10^{-5}$   &   $ 7.57$   &   $ 0.70$\\
Cl   &   $ 1.67$   &   $3.80\times 10^{-4}$   &   $1.49\times 10^{-5}$   &   $ 7.18$   &   $ 1.94$\\
Ar   &   $ 0.45$   &   $2.08\times 10^{-4}$   &   $7.23\times 10^{-6}$   &   $ 6.87$   &   $ 0.72$\\
Ca   &   $ 0.67$   &   $2.84\times 10^{-4}$   &   $9.82\times 10^{-6}$   &   $ 7.00$   &   $ 0.92$\\
Sc   &$\le 1.79$   &$\le2.93\times 10^{-6}$   &$\le1.00\times 10^{-7}$   &$\le 4.96$   &$\le 2.11$\\
Ti   &$\le 1.02$   &$\le3.12\times 10^{-5}$   &$\le1.00\times 10^{-6}$   &$\le 5.96$   &$\le 1.31$\\
V    &$\le 2.06$   &$\le3.32\times 10^{-5}$   &$\le1.00\times 10^{-6}$   &$\le 5.96$   &$\le 2.33$\\
Cr   &$\le 1.33$   &$\le3.39\times 10^{-4}$   &$\le1.00\times 10^{-5}$   &$\le 6.96$   &$\le 1.62$\\
Mn   &$\le 0.53$   &$\le3.58\times 10^{-5}$   &$\le1.00\times 10^{-6}$   &$\le 5.96$   &$\le 0.83$\\
Fe   &   $-0.24$   &   $6.92\times 10^{-4}$   &   $1.72\times 10^{-5}$   &   $ 7.24$   &   $ 0.00$\\
Co   &$\le 2.02$   &$\le3.84\times 10^{-4}$   &$\le1.00\times 10^{-5}$   &$\le 6.96$   &$\le 2.27$\\
Ni   &$\le 0.75$   &$\le3.82\times 10^{-4}$   &$\le1.00\times 10^{-5}$   &$\le 6.96$   &$\le 1.04$\\
\hline
\end{tabular}
\newline
\textbf{Notes. }
{
$^{(a)}${\citet{baillerjonesetal2018}},
$^{(b)}${interpolated from evolutionary tracks of \citet{dormanetal1993}},
$^{(c)}${\citet{2011ApJ...737..103SchlaflyFinkbeiner}},
$^{(d)}${\citet{bianchietal2017}},
$^{(e)}${\citet{2012yCat.1322....0ZachariasUCAC4}},
$^{(f)}${\citet{kupfer2015A&A...576A..44K}},
$^{(g)}${\citet{cutri2003}}.
}}
\end{table*}

\begin{figure*}
  \resizebox{\hsize}{!}{\includegraphics{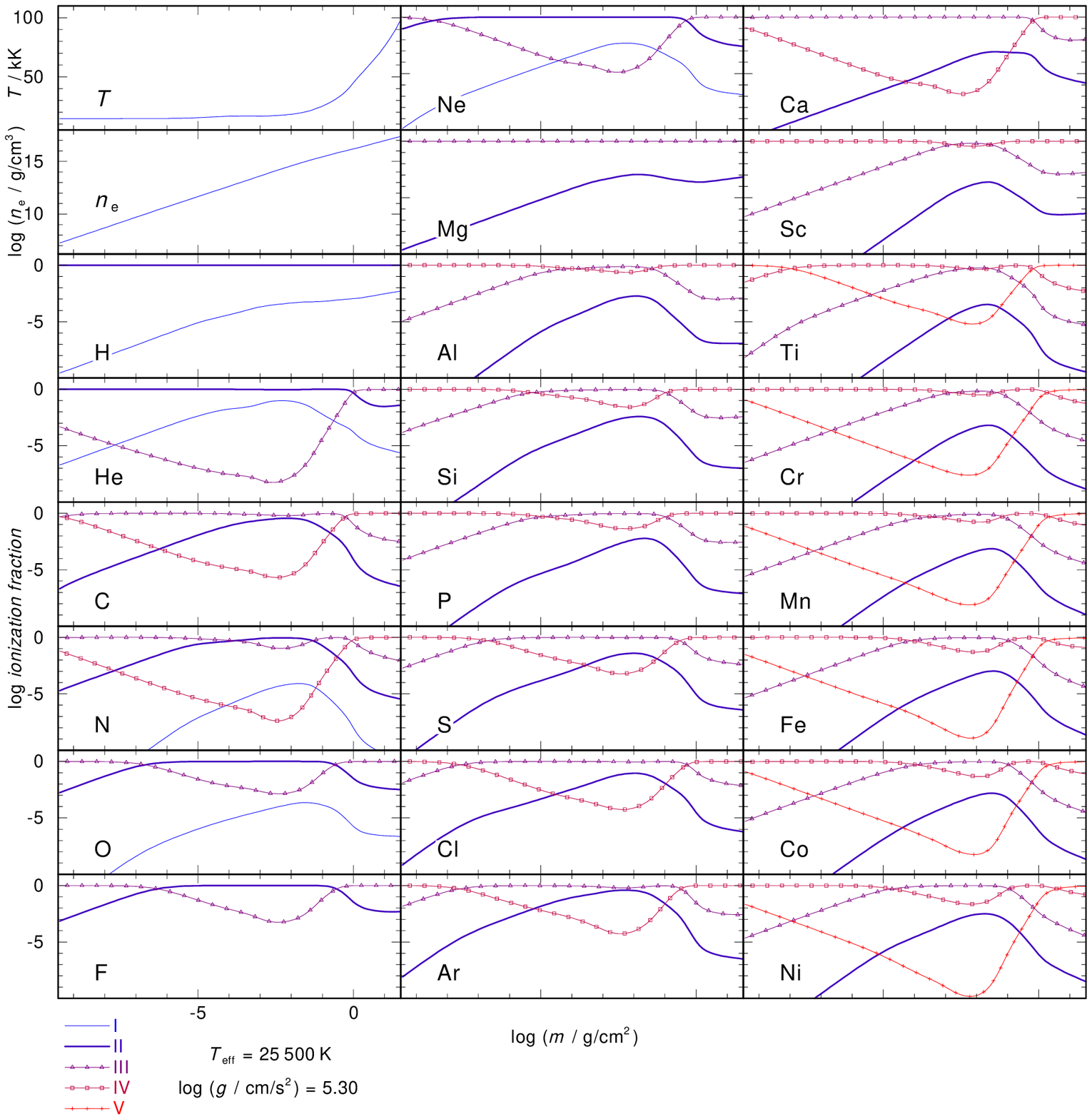}} 
   \caption[]{Temperature and {electron} density structure and ionization fractions of all ions which are considered in our final model for \cpd.
             } 
   \label{fig:ionfrac}
\end{figure*}

\begin{figure*}
  \resizebox{\hsize}{!}{\includegraphics{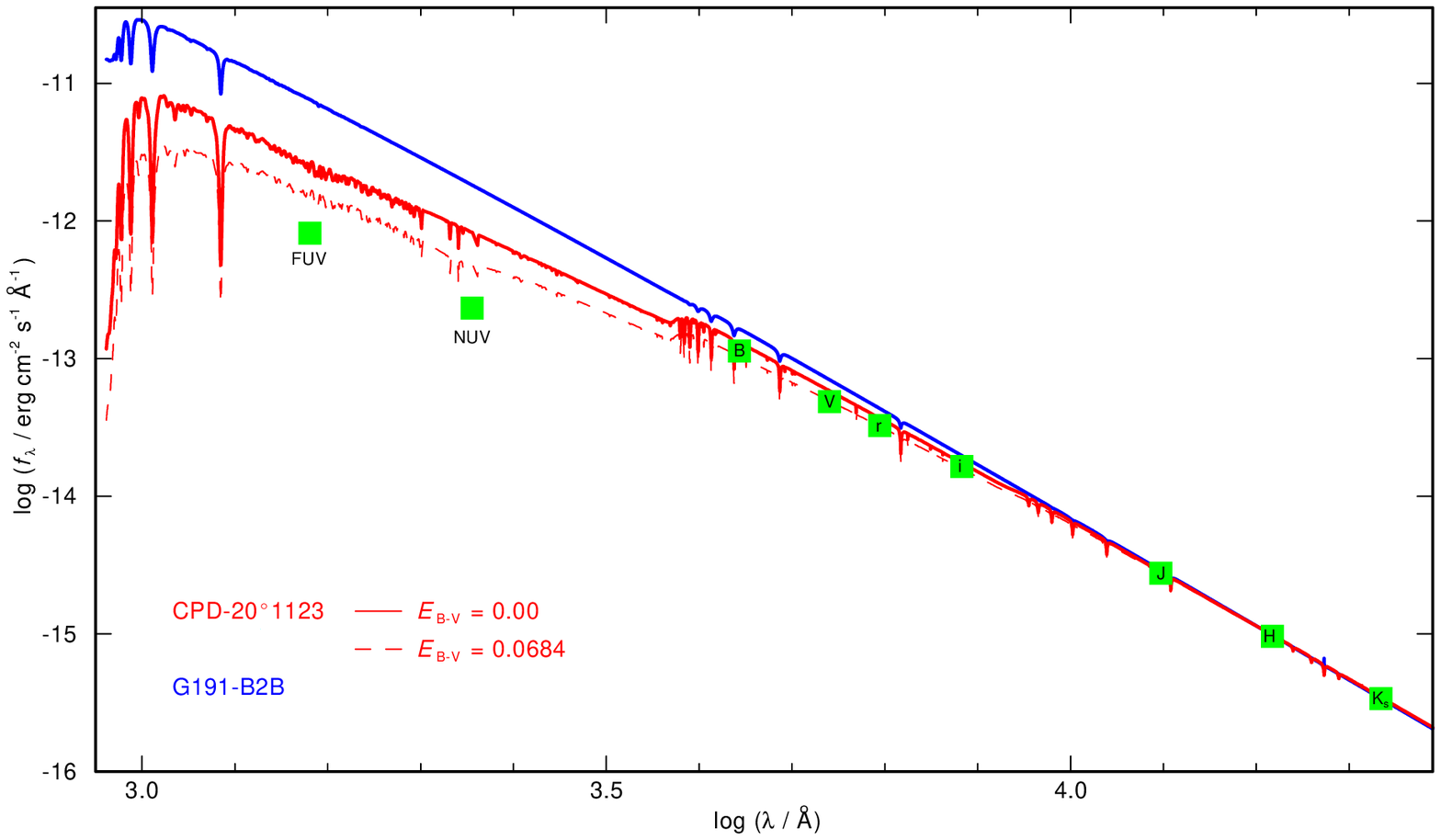}} 
   \caption[]{{Synthetic spectra of our best model of \cpd without reddening (thick, red line) and with interstellar reddening with \ebvw{0.0684} (dashed, red) compared with the SED of the DA WD G191$-$B2B {(blue)}. The model fluxes are normalized to the 2MASS $K_s$ magnitude \citep[$\lambda = 21590$\,{\AA},][]{cutri2003}.  
               $B$, $r$, and $i$ magnitudes from \citet[][]{2012yCat.1322....0ZachariasUCAC4}, the $V$ magnitude from \citet{kupfer2015A&A...576A..44K}, and the GALEX FUV and NUV magnitudes from \citet{bianchietal2017} were added.}} 
   \label{fig:ebv}
\end{figure*}

\begin{figure*}
  \resizebox{\hsize}{!}{\includegraphics{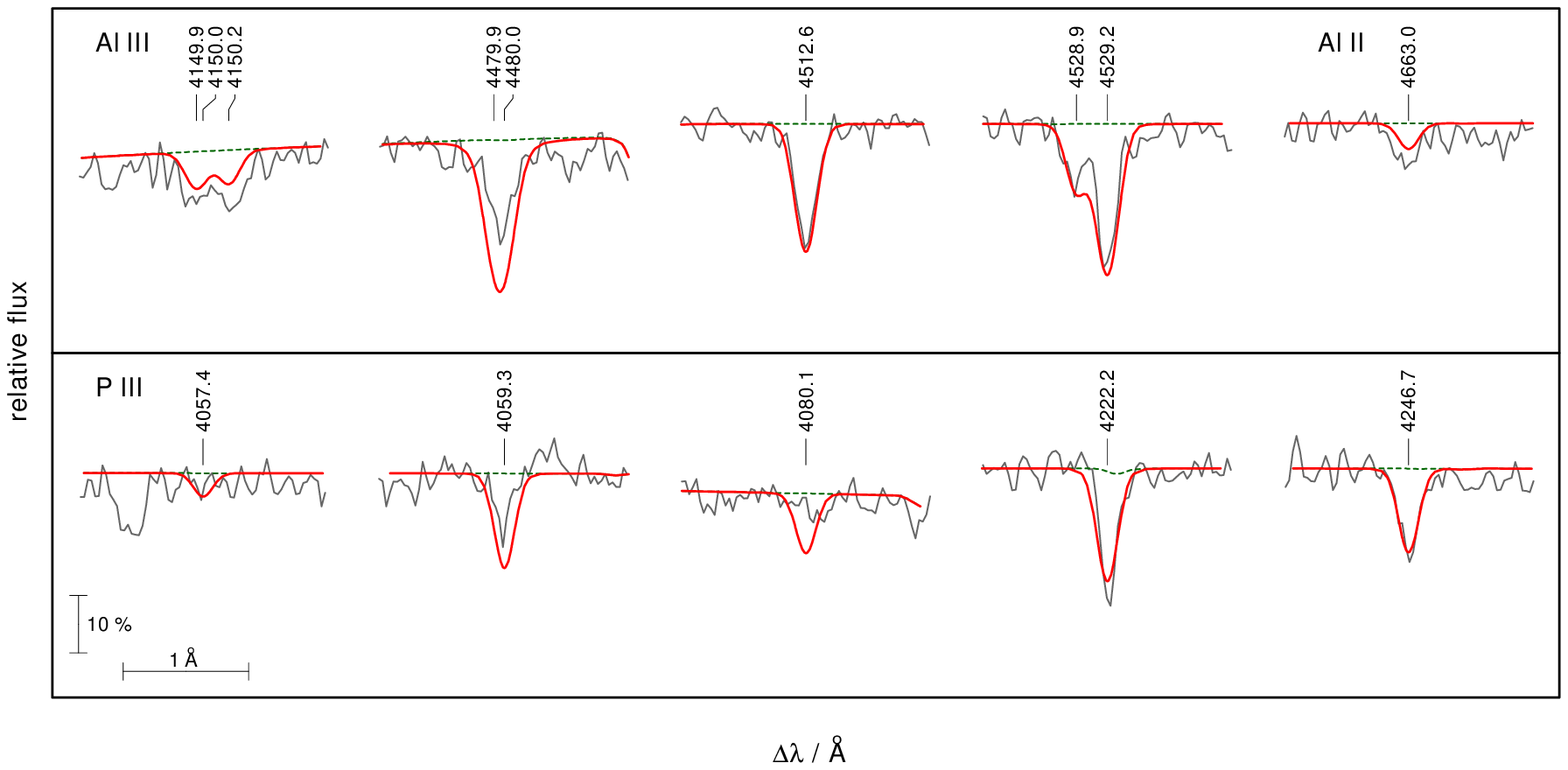}} 
   \caption[]{Synthetic spectra calculated with \loggw{5.30}, $\Teff = 25\,500$, and final abundances from \ta{tab:finab}, compared with the UCLES observations of lines of \Ion{Al}{2}, \Ion{Al}{3}, and \Ion{P}{3}.} 
   \label{fig:alplines}
\end{figure*}

\begin{figure*}
  \resizebox{\hsize}{!}{\includegraphics{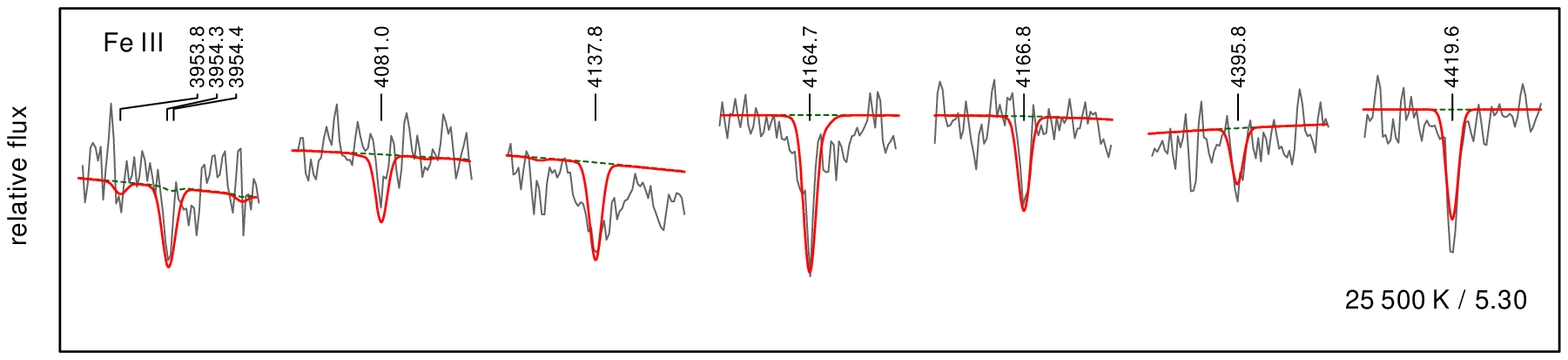}} 
   \caption[]{{Synthetic spectra calculated with \loggw{5.30}, $\Teff = 25\,500$, and final abundances from \ta{tab:finab}, compared with the UCLES observations of lines of \Ion{Fe}{3}.}} 
   \label{fig:felines}
\end{figure*}

%%%%%%%%%%%%%%%%%%%%%%%%%%%%%%%%%%%%%%%%%%%%%%%%%%

% Don't change these lines
\bsp	% typesetting comment
\label{lastpage}
\end{document}